\documentclass{llncs}
\usepackage{amssymb, amsmath}
\usepackage[utf8]{inputenc}
\usepackage[T1]{fontenc}
\usepackage{unicode}
\usepackage{tikz}
\usepackage{hyperref}
\usepackage{multirow}
\usepackage{bigdelim}
\usepackage[ruled, vlined, linesnumbered]{algorithm2e}
\SetEndCharOfAlgoLine{}
\SetKwInput{Params}{Public parameters}
\SetKwInput{Key}{Shared key}
\SetKwInput{Input}{Input}
\SetKwInput{Output}{Output}
\SetKwBlock{Alice}{Alice}{}
\SetKwBlock{Bob}{Bob}{}
\SetKw{Samples}{samples}
\SetKw{Sends}{sends}
\SetKw{Sends}{sends}
\SetKwProg{Function}{function}{}{}
\SetKwFunction{KeyGen}{KeyGen}
\SetKwFunction{DH}{DH}

\newcommand{\F}{\mathbb{F}}
\newcommand{\Fbar}{\overline{\mathbb{F}}}

\newcommand{\Z}{\mathbb{Z}}
\newcommand{\Cl}{\mathcal{C}}
\newcommand{\Graph}{\mathcal{G}}
\renewcommand{\O}{\mathcal{O}}
\newcommand{\softO}{\tilde{O}}

\newcommand{\set}[1]{\left\{#1\right\}}
\newcommand{\suchthat}{\,\middle\vert\,}
\newcommand{\algstyle}[1]{\textsc{#1}}

\renewcommand{\frak}{\mathfrak}
\newcommand{\rand}[1]{\overset{#1}{∈}}
\newcommand{\uni}{\rand{R}}
\newcommand{\Adv}[2][]{\mathsf{Adv}^{#1}_{\text{\rm #2}}}

\DeclareMathOperator{\End}{End}

\DeclareMathOperator{\Ell}{Ell}
\DeclareMathOperator{\poly}{poly}
\DeclareMathOperator{\Proba}{Pr}

\begin{document}

\title{Towards practical key exchange\\ from ordinary isogeny graphs}
\author{
 Luca De Feo\inst{1,3}\orcidID{0000-0002-9321-0773} \and
 Jean Kieffer\inst{2,3,4} \and
 Benjamin Smith\inst{3}
}
\institute{
 Université Paris Saclay, UVSQ, LMV, Versailles, France
 \\
 \email{luca.de-feo@uvsq.fr}
 \and
 École Normale Supérieure, Paris, France
 \\
 \email{jean.kieffer.14@normalesup.org}
 \and
 Inria and École polytechnique, Université Paris Saclay, Palaiseau, France
 \\
 \email{smith@lix.polytechnique.fr}
 \and
 IMB - Institut de Mathématiques de Bordeaux, Inria Bordeaux - Sud-Ouest, Talence, France
}

\maketitle

\begin{abstract}
    We revisit the ordinary isogeny-graph based cryptosystems
    of Couveignes and Rostovtsev--Stolbunov,
    long dismissed as impractical.
    We give algorithmic improvements that accelerate key exchange
    in this framework,
    and explore the problem of generating suitable system parameters
    for contemporary pre- and post-quantum security that take
    advantage of these new algorithms.
    We also prove the session-key security of this key exchange
    in the Canetti--Krawczyk model,
    and the IND-CPA security of the related public-key encryption scheme,
    under reasonable assumptions on the hardness of computing isogeny walks.
    Our systems admit efficient key-validation techniques that
    yield CCA-secure encryption,
    thus providing an important step towards
    efficient post-quantum non-interactive key exchange (NIKE).

\end{abstract}

\section{Introduction}
\label{sec:introduction}

Isogeny-based protocols
form one of the youngest and least-explored
families of post-quantum candidate cryptosystems.
The best-known isogeny-based protocol
is Jao and De Feo's SIDH key exchange~\cite{jao+defeo2011},
from which the NIST candidate key-encapsulation mechanism
SIKE was derived~\cite{SIKE,NIST2016}.
SIDH was itself inspired by earlier
key-exchange constructions
by Couveignes~\cite{cryptoeprint:2006:291}
and Rostovtsev and
Stolbunov~\cite{rostovtsev+stolbunov06,stolbunov-red,Stol},
which were widely considered unwieldy and impractical.

Indeed, the origins of isogeny-based cryptography can be traced back to
Couveignes' ``Hard Homogeneous Spaces'' manuscript,
that went unpublished for ten years before appearing
in~\cite{cryptoeprint:2006:291}. 
A \emph{principal homogeneous space} (PHS) for a group $G$ is a set $X$ with an
action of $G$ on $X$ such that for any $x,x'\in X$, there is a
unique $g\in G$ such that $g\cdot x = x'$. Equivalently,
the map $φ_x: g\mapsto g\cdot x$
is a bijection between $G$ and $X$ for any $x\in X$.
Couveignes defines a \emph{hard homogeneous space} (HHS) 
to be a PHS where the action of $G$ on $X$ is efficiently computable, 
but inverting the isomorphism $φ_x$ is computationally hard for any~$x$.

\begin{algorithm}
    \caption{Key generation for cryptosystems in an HHS $X$ for a
    group $G$, with a fixed ``base point'' $x_0$ in $X$.}
    \label{alg:HHS-KeyGen}
    \KwIn{()}
    \KwOut{A private-public keypair $(g,x)\in G\times X$
    s.t. $x = g\cdot x_0$}
    \Function{\KeyGen{}}{
        $g \gets \algstyle{Random}(G)$
        \tcp*{$g$ is sampled uniformly at random from $G$}
        $x \gets g\cdot x_0$
        \;
        \Return{$(g,x)$}
    }
\end{algorithm}

Any HHS $X$ for an \emph{abelian} group $G$ can be used to construct a
key exchange based on the hardness of inverting $\varphi_x$,
as shown in Algorithms~\ref{alg:HHS-KeyGen} and~\ref{alg:HHS-DH}.
If Alice and Bob have keypairs $(g_A,x_A)$
and $(g_B,x_B)$, respectively,
then the commutativity of $G$
lets them derive a shared secret
\[
    \DH(g_A,x_B) 
    = g_A\cdot(g_B\cdot x_0)
    = g_B\cdot(g_A\cdot x_0)
    = \DH(g_B,x_A)
    \,.
\]
The analogy with classic group-based Diffie--Hellman is evident.

\begin{algorithm}
    \caption{Diffie--Hellman in an HHS $X$ for a group $G$}
    \label{alg:HHS-DH}
    \KwIn{A private key $g_A\in G$ and a public key $x_B\in X$,
    each generated by calls to \KeyGen}
    \KwOut{A shared secret value $k\in X$}
    \Function{\DH{$g_A$,$x_B$}}{
        $k \gets g_A\cdot x_B$
        \;
        \Return{$k$}
    }
\end{algorithm}

For example, if $X=\langle{x}\rangle$ is cyclic of order $p$ 
and $G=(\Z/p\Z)^*$ acts on $X\setminus\{1\}$ by
$g·x=x^g$, then inverting $φ_x$ is the discrete logarithm problem (DLP) in $X$.
But inverting $φ_x$ for other homogeneous spaces may not be related
to any DLP,
and might resist attacks based on Shor's quantum algorithm.
Similar ideas have occasionally appeared in the
literature in different forms~\cite{10.1007/3-540-44598-6_10,monico2007}.

Couveignes viewed HHS chiefly as a general framework encompassing
various Diffie--Hellman-like systems. Nevertheless, he
suggested using a specific HHS based on the theory of complex
multiplication of elliptic curves, in a sense generalizing 
Buchmann and Williams'
class-group-based Diffie--Hellman key exchange~\cite{Buchmann1988}.
Independently, Rostovtsev
and Stolbunov proposed in~\cite{rostovtsev+stolbunov06} a public key
encryption scheme based on the same HHS. Later, Stolbunov~\cite{Stol}
derived more protocols from this, including an interactive
key exchange scheme similar to Algorithm~\ref{alg:HHS-DH}.  Rostovtsev
and Stolbunov's proposal deviates from the HHS paradigm in
the way random elements of $G$ are sampled, as we will explain in
\S\ref{sec:keyex}. This makes the primitive less flexible, but
also more practical.

Rostovtsev and Stolbunov advertised their cryptosystems as potential
post-quantum candidates, leading Childs, Jao and Soukharev to introduce
the first subexponential quantum algorithm capable of breaking
them~\cite{childs2014constructing}. 
Hence, being already slow enough to
be impractical in a classical security setting, their
primitive appeared even more impractical in a quantum
security setting. 

But the Couveignes--Rostovtsev--Stolbunov primitive (CRS)
has some important advantages over SIDH which make it worth pursuing.
Unlike SIDH, 
CRS offers efficient and safe public key validation,
making it suitable for non-interactive key exchange (NIKE).
Further, CRS does not suffer from 
some of the potential cryptographic weaknesses that SIDH has,
such as short paths and the publication of image points.

This paper aims to improve and modernize the CRS construction, 
borrowing techniques from SIDH and point-counting algorithms,
to the point of making it usable in a post-quantum setting.  
Our main contributions
are in \S\S\ref{sec:keyex}--\ref{sec:initcurve}, where we present
a new, more efficient way of computing the CRS group action, and in
\S\ref{sec:sec}, where we give precise classic and quantum
security estimates, formalize hardness assumptions, and sketch
security proofs in stronger models than those previously
considered. 
In \S\ref{sec:exp} we present a
proof-of-concept implementation and measure its performance.
While the final result is far from competitive, we believe it
constitutes progress towards a valid isogeny-based
alternative to SIDH.

\paragraph{CSIDH.}
While preparing this paper we were informed of
recent work by Castryck, Lange, Martindale, Panny, and Renes,
introducing CSIDH, 
an efficient post-quantum primitive based on CRS~\cite{csidh}.
Their work builds upon the ideas presented in
\S\S\ref{sec:keyex}--\ref{sec:initcurve}, using them in a
different homogeneous space where they apply effortlessly.  Their
breakthrough confirms that, if anything, our techniques were a
fundamental step towards the first practical post-quantum
non-interactive key exchange protocol.

\paragraph{Side channel awareness.}
The algorithms we present here are not intended to provide any protection
against basic side-channel attacks.  
Uniform and constant-time algorithms for arbitrary-degree isogeny computations
are an interesting open problem,
but they are beyond the scope of this work.

\paragraph{Acknowledgments.}
We would like to thank Wouter Castryck, Tanja Lange, Chloe Martindale,
Lorenz Panny, and Joost Renes for sharing a draft of their paper with
us, and Alexandre Gélin and François Morain for fruitful discussions.
De Feo acknowledges the support of the French \emph{Programme
  d'Investissements d'Avenir} under the national project RISQ n°
P141580-3069086/DOS0044212.

\section{Isogenies and complex multiplication}
\label{sec:math}

We begin by recalling some basic facts on isogenies of elliptic curves 
over finite fields. For an in-depth introduction to these concepts, we
refer the reader to~\cite{silverman:elliptic}. For a general
overview of isogenies and their use in cryptography, we
suggest~\cite{defeo2017isogenybased}.

\subsection{Isogenies between elliptic curves}
\label{sec:isogeny}

In what follows $\F_q$ is a finite field of characteristic $p$ with
$q$ elements, and $\Fbar_q$ is its algebraic closure. Let $E$ and $E'$
be elliptic curves defined over $\F_q$. 
A homomorphism $ϕ:E→E'$ is an
algebraic map sending $0_E$ to $0_{E'}$;
it induces a group homomomorphism from
$E(\Fbar_q)$ to $E'(\Fbar_q)$~\cite[III.4]{silverman:elliptic}.
An \emph{endomorphism} is a homomorphism from a curve to itself.
The endomorphisms of $E$ form a ring $\End(E)$,
with the group law on $E$ for addition
and composition for multiplication.
The simplest examples of endomorphisms
are the scalar multiplications $[m]$
(mapping $P$ to the sum of $m$ copies of $P$)
and the \emph{Frobenius} endomorphism
\begin{align*}
  π : E &\longrightarrow E \,, \\
  (x,y) &\longmapsto (x^q,y^q) \,.
\end{align*}
As an element of $\End(E)$, Frobenius satisfies a quadratic equation
$π^2 + q = tπ$.  The integer $t$ (the \emph{trace})
fully determines the order of $E$ as $\#E(\F_q)=q+1-t$. A curve is
called \emph{supersingular} if $p$ divides $t$, \emph{ordinary}
otherwise.

An \emph{isogeny} is a non-zero homomorphism of elliptic curves.
The
degree of an isogeny is its degree as an algebraic map,
so for example the Frobenius endomorphism $\pi$ has degree $q$,
and the scalar multiplication $[m]$ has degree $m^2$.
Isogenies of degree $ℓ$ are called $ℓ$-isogenies.
The kernel $\ker ϕ$ of $\phi$
is the subgroup of $E(\Fbar_q)$ that is
mapped to $0_{E'}$. 
An isogeny $ϕ$ is \emph{cyclic} 
if $\ker ϕ$ is a cyclic group.

An \emph{isomorphism} is an isogeny of degree~\(1\).
An \emph{isomorphism class} of elliptic curves is
fully determined by their common \emph{$j$-invariant} in $\Fbar_q$. 
If any curve in the isomorphism class is defined
over $\F_q$, then its $j$-invariant is in $\F_q$.

Any isogeny can be factored as a composition of a \emph{separable} and
a \emph{purely inseparable} isogeny. \emph{Purely inseparable}
isogenies have trivial kernel, and degree a power of $p$.
\emph{Separable} isogenies include all
isogenies of degree coprime to $p$.
Up to isomorphism, separable isogenies
are in one-to-one correspondence with their kernels:
for any finite subgroup $G⊂E$ of order $ℓ$ there is 
an elliptic curve $E/G$ and an $\ell$-isogeny $\phi: E \to E/G$
such that $\ker \phi = G$,
and the curve and isogeny are unique up to isomorphism.
In particular, if $\phi$ is separable then $\deg ϕ=\#\ker ϕ$.
It is convenient to encode $\ker\phi$ as
the polynomial whose roots are the $x$-coordinates of the points
in $\ker\phi$, called the \emph{kernel polynomial} of $\phi$.

For any $ℓ$-isogeny $ϕ:E→E'$, there is a unique $ℓ$-isogeny
$\hat{ϕ}:E'→E$ such that $ϕ∘\hat{ϕ} = [\ell]$ on $E'$
and $\hat{ϕ}∘ϕ = [\ell]$ on $E$.
We call $\hat{ϕ}$ the \emph{dual} of $ϕ$. This
shows that being \emph{$\ell$-isogenous} is a symmetric
relation, and that being isogenous is an equivalence relation.
 Further, a theorem of Tate states that two curves are
isogenous over $\F_q$ if and only if they have the same number of
points over $\F_q$.

\subsection{Isogeny graphs}
\label{sec:isogeny-graphs}

Isogeny-based cryptosystems are based on \emph{isogeny graphs}.
These are
(multi)-graphs whose vertices are
elliptic curves up to isomorphism, and whose edges are isogenies
between them (again up to isomorphism).
The use of isogeny graphs for algorithmic applications 
goes back to Mestre and Oesterlé~\cite{Mestre},
followed notably by Kohel~\cite{kohel},
and has been continued by many
authors~\cite{Gal,fouquet+morain02,GHS,MiretMSTV06,jao+miller+venkatesan09}.

We write $E[ℓ]$ for the subgroup of $ℓ$-torsion points of
$E(\Fbar_q)$.  If $ℓ$ is coprime to $p$, then $E[ℓ]$ is isomorphic to
$(ℤ/ℓℤ)^2$.  Furthermore, if $ℓ$ is prime then $E[ℓ]$ contains exactly
$ℓ+1$ cyclic subgroups of order $ℓ$; it follows that, over $\Fbar_q$,
there are exactly $ℓ+1$ distinct (non-isomorphic) separable $ℓ$-isogenies 
from $E$ to other curves.
Generically, a connected component of the $\ell$-isogeny graph 
over $\Fbar_q$ will be an infinite $(ℓ+1)$-regular
graph (a notable exception is the finite connected component of
\emph{supersingular} curves, used in SIDH and related protocols).

We now restrict to isogenies defined over $\F_q$.
If $E$ and $E'$ are elliptic curves over $\F_q$,
then an isogeny $ϕ:E→E'$ is defined over $\F_q$
(up to a twist of $E'$)
if and only if the Frobenius endomorphism $\pi$ on $E$ stabilizes $\ker ϕ$.
We emphasize that the points in $\ker\phi$ need not
be defined over $\F_q$ themselves.

For the vertices of the $\Fbar_q$-isogeny graph
we use $j$-invariants,
which classify elliptic curves up to
$\Fbar_q$-isomorphism;
but in the sequel we want to work up to $\F_q$-isomorphism,
a stronger equivalence.
If $E$ and $\tilde{E}$ are not $\F_q$-isomorphic
but $j(E) = j(\tilde{E})$,
then $\tilde{E}$ is the \emph{quadratic twist} of $E$
(which is defined and unique up to $\F_q$-isomorphism).\footnote{%
    There is a slight technicality here for $j$-invariants $0$ and $1728$,
    where non-quadratic twists may exist.
    We ignore these special cases
    because these curves never appear in our cryptosystem:
    the class groups of their endomorphism rings are trivial,
    and keyspaces of size 1 are of limited utility in cryptography.
}
When $E$ is ordinary,
its quadratic twist has a different cardinality
(if $\#E(\F_q) = q + 1 - t$, then $\#\tilde{E}(\F_q) = q + 1 + t$),
so $E$ and $\tilde{E}$ are in different components of the isogeny graph.
But every $\F_q$-isogeny $\phi: E \to E'$ 
corresponds to an $\F_q$-isogeny $\tilde{\phi}: \tilde{E} \to \tilde{E}'$
of the same degree between the quadratic twists.
The component of the $\F_q$-isogeny graph containing an ordinary curve 
and the component containing its twist are thus isomorphic;
we are therefore justified in identifying them,
using $j$-invariants in $\F_q$ for vertices in the
$\F_q$-graph.\footnote{%
    The situation is much more complicated for supersingular graphs,
    because the curve and its twist are in the same component
    of the graph; see~\cite[\S2]{DelfsG16} for details.
}
This is not just a mathematical convenience:
we will see in \S\ref{sec:keyex} below 
that switching between a curve and its twist
often allows a useful optimization in isogeny computations.

If an isogeny $ϕ$ is defined over $\F_q$ \emph{and cyclic},
then $π$ acts like a scalar on the points of $\ker ϕ$. 
Thus, for any prime $ℓ≠p$, the number of outgoing $ℓ$-isogenies from $E$ 
defined over $\F_q$ can be
completely understood by looking at how $π$ acts on $E[ℓ]$. Since $E[ℓ]$
is a $ℤ/ℓℤ$-module of rank $2$, the action of $π$ is represented by a
$2×2$ matrix with entries in $ℤ/ℓℤ$ and characteristic polynomial
$X^2-tX+q\mod ℓ$. We then have four possibilities:
\begin{itemize}
\item[(0)] $π$ has no eigenvalues in $ℤ/ℓℤ$, i.e.\ $X^2-tX+q$ is
  irreducible modulo $ℓ$; then $E$ has no $ℓ$-isogenies.
\item[(1.1)] $π$ has one eigenvalue of (geometric) multiplicity one,
  i.e.\ it is conjugate to a non-diagonal matrix
  $\left(\begin{smallmatrix}λ&*\\0&λ\end{smallmatrix}\right)$; then
  there is one $ℓ$-isogeny from $E$.
\item[(1.2)] $π$ has one eigenvalue of multiplicity two, i.e.\ it acts
  like a scalar matrix
  $\left(\begin{smallmatrix}λ&0\\0&λ\end{smallmatrix}\right)$; then
  there are $ℓ+1$ isogenies of degree $ℓ$ from $E$.
\item[(2)] $π$ has two distinct eigenvalues, i.e.\ it is conjugate to a
  diagonal matrix
  $\left(\begin{smallmatrix}λ&0\\0&μ\end{smallmatrix}\right)$
	with $\lambda\neq\mu$; then
  there are two $\ell$-isogenies from $E$.
\end{itemize}

The primes $\ell$ in Case~(2)
are called \emph{Elkies primes} for $E$;
these are the primes of most interest to us.
Cases~(1.x) are only possible if $ℓ$ divides $Δ_π = t^2-4q$,
the discriminant of the characteristic equation of $π$;
for ordinary curves $Δ_π≠0$, so only a finite number
of $ℓ$ will fall in these cases, and they will be mostly
irrelevant to our cryptosystem.
We do not use any $\ell$ in Case~(0).

Since all curves in
the same isogeny class over $\F_q$ have the same number of points,
they also have the same trace $t$ and discriminant $Δ_π$.
It follows that if $\ell$ is Elkies for some $E$ in $\Ell_q(\O)$,
then it is Elkies for every curve in $\Ell_q(\O)$.

Hence, if $ℓ$ is an Elkies prime for a curve $E$,
then the connected component of $E$ in the $\ell$-isogeny graph 
is a finite $2$-regular graph---that is, a cycle. 
In the next subsection we describe a group action on this cycle,
and determine its size.

\subsection{Complex multiplication}

In this subsection we focus exclusively on ordinary elliptic curves. 
If $E$ is an ordinary curve with Frobenius $π$,
then $\End(E)$ is isomorphic to an
\emph{order}\footnote{%
    An \emph{order} is a subring which is a $ℤ$-module of rank $2$.
} in the quadratic imaginary field
$ℚ(\sqrt{Δ_π})$ (see~\cite[III.9]{silverman:elliptic}).
A curve whose endomorphism ring is isomorphic to an order $\O$ is said to
have \emph{complex multiplication} by $\O$.
For a detailed treatment of the theory of complex multiplication,
see~\cite{lang1987elliptic,silverman:advanced}.

The ring of integers $\O_K$ of $K=ℚ(\sqrt{Δ_π})$ is its
\emph{maximal order}: it contains any other order of $K$.  Hence
$ℤ[π]⊂\End(E)⊂\O_K$, and there is only a finite number of possible
choices for $\End(E)$. If we write $Δ_π=d^2Δ_K$, where $Δ_K$ is the
discriminant\footnote{%
    $Δ_K$ is 
    a \emph{fundamental discriminant}:
    $\Delta_K\equiv0,1\pmod 4$, and $Δ_K$ or $\frac{Δ_K}{4}$ is squarefree.
} 
of $\O_K$, then the index $[\O_K:\End(E)]$ must divide $d=[\O_K:ℤ[π]]$.

It turns out that isogenies allow us to navigate the various
orders. If $ϕ:E→E'$ is an $\ell$-isogeny, then one of the following
holds~\cite[Prop.~21]{kohel}:
\begin{itemize}
\item $\End(E) = \End(E')$, and then $ϕ$ is said to be
  \emph{horizontal};
\item $[\End(E):\End(E')] = ℓ$, and then $ϕ$ is said to be
  \emph{descending};
\item $[\End(E'):\End(E)] = ℓ$, and then $ϕ$ is said to be
  \emph{ascending}.
\end{itemize}
Notice that the last two cases can only happen if $ℓ$ divides
$d^2=Δ_π/Δ_K$, and thus correspond to Cases (1.x) in the previous
subsection.
If $ℓ$ does not divide $Δ_π$, then $ϕ$ is necessarily horizontal.

We now present a group action on the set of all curves 
up to isomorphism having complex
multiplication by a fixed order $\O$; the key exchange protocol of
\S\ref{sec:keyex} will be built on this action. Let $\frak a$ be
an invertible ideal in $\End(E)≃\O$ of norm prime to $p$, and define the
\emph{${\frak a}$-torsion} subgroup of $E$ as
\[
E[\frak a] = \set{P\in E(\Fbar_q) \suchthat σ(P) = 0\ 
\text{ for all }σ\in\frak a}.
\]
This subgroup is the kernel of a separable isogeny $\phi_{\frak a}$. \footnote{In fact, one can define $\phi_{\frak a}$ for any invertible ideal $\frak a$, but it is not always separable.}
The codomain $E/E[\frak a]$ of $\phi_{\frak a}$ is well-defined up to isomorphism
and will be denoted $\frak a\cdot E$.
The isogeny $\phi_{\frak a}$ is
always horizontal---that is, $\End(\frak a \cdot E) = \End(E)$---and its
degree is the \emph{norm} of $\frak a$ as an ideal of $\End(E)$.

Let $\Ell_q(\O)$ be the set of isomorphism classes over $\Fbar_q$
of curves with complex multiplication by $\O$, and assume it is non-empty. 
The construction above gives rise
to an action of the group of fractional ideals of $\O$ on $\Ell_q(\O)$.
Furthermore, the principal ideals act trivially 
(the corresponding isogenies are endomorphisms), 
so this action induces an action of the \emph{ideal
class group} $\Cl(\O)$ on $\Ell_q(\O)$.

The main theorem of complex
multiplication states that this action is \emph{simply transitive}. In
other terms, $\Ell_q(\O)$ is a PHS
under the group $\Cl(\O)$: if we fix a curve $E$ as base point,
then we have a bijection
\[
\begin{aligned}
\Cl(\O) &\longrightarrow \Ell_q(\O) \\
\text{Ideal class of }\frak a &\longmapsto \text{Isomorphism class of }\frak a\cdot E.
\end{aligned}
\]
The order of $\Cl(\O)$ is called the \emph{class number} of $\O$, and
denoted by $h(\O)$. An immediate consequence of the theorem is that
$\#\Ell_q(\O)=h(\O)$.

As before, we work with $\F_q$-isomorphism classes.
Then $\Ell_q(\O)$ decomposes into two isomorphic PHSes under $\Cl(\O)$,
each containing the quadratic twists of the curves in the other. 
We choose one of these
two components, that we will also denote $\Ell_q(\O)$ in the sequel.
(The choice is equivalent to a choice of isomorphism $\End(E) \cong \O$,
and thus to a choice of sign on the image of~\(\pi\) in $\O$.)

Now let $ℓ$ be an Elkies prime for $E\in\Ell_q(\O)$. So far, we have seen that the
connected component of $E$ in the $ℓ$-isogeny graph is a cycle of
horizontal isogenies. Complex multiplication tells us more. The
restriction of the Frobenius to $E[ℓ]$ has two eigenvalues $λ≠μ$, to
which we associate the prime ideals $\frak a=(π-λ,ℓ)$ and
$\hat{\frak a}=(π-μ,ℓ)$, both of norm $ℓ$. We see then that
$E[\frak a]$ is the eigenspace of $λ$, defining an isogeny
$ϕ_{\frak{a}}$ of degree $ℓ$. Furthermore
$\frak a\hat{\frak a} = \hat{\frak a}\frak a = (ℓ)$, implying that
$\frak a$ and $\hat{\frak a}$ are the inverse of one another in
$\Cl(\O)$, thus the isogeny $ϕ_{\hat{\frak a}}:\frak a·E→E$ of
kernel $(\frak a·E)[\hat{\frak a}]$ is the dual of $ϕ_{\frak a}$ (up
to isomorphism). 

The eigenvalues $λ$ and $μ$ define opposite directions on the
$\ell$-isogeny cycle,
independent of the starting curve,
as shown in Figure~\ref{fig:cycle}.  
The size of the cycle is the order of $(π-λ,ℓ)$ in $\Cl(\O)$,
thus partitioning $\Ell_q(\O)$ into cycles of equal size.

\begin{figure}[t]
  \begin{minipage}{0.45\textwidth}
    \centering
    \begin{tikzpicture}
      \def\crater{7}
      \foreach \i in {1,...,\crater} {
        \begin{scope}[shorten >=0.1cm,->]
          \draw[blue!60!black] (360/\crater*\i : 1.95cm) -- (360/\crater*\i+360/\crater : 1.95cm);
          \draw[blue!60!white] (360/\crater*\i+360/\crater : 2.05cm) -- (360/\crater*\i : 2.05cm);
        \end{scope}
        \draw[blue!60!black] (360/\crater*\i+180/\crater:1.6cm) node {\small$λ$};
        \draw[blue!60!white] (360/\crater*\i+180/\crater:2cm) node {\small$μ$};
      }
      \foreach \i in {1,...,\crater} {
        \draw[fill] (360/\crater*\i:2cm) circle (2pt);
      }
    \end{tikzpicture}
    \caption{An isogeny cycle for an Elkies prime $ℓ$, with edge directions
      associated with the Frobenius eigenvalues $λ$ and $μ$.}
    \label{fig:cycle}
  \end{minipage}
  \hfill
  \begin{minipage}{0.45\textwidth}
    \centering
    \begin{tikzpicture}
      \def\crater{12}
      \def\jumpa{-8}
      \def\jumpb{9}
      \def\diam{2cm}

      \foreach \i in {1,...,\crater} {
        \draw[blue] (360/\crater*\i : \diam) to[bend right] (360/\crater*\i+360/\crater : \diam);
        \draw[red] (360/\crater*\i : \diam) to[bend right] (360/\crater*\i+\jumpa*360/\crater : \diam);
        \draw[green] (360/\crater*\i : \diam) to[bend right=50] (360/\crater*\i+\jumpb*360/\crater : \diam);
      }
      \foreach \i in {1,...,\crater} {
        \pgfmathparse{int(mod(2^\i,13))}
        \let\exp\pgfmathresult
        \draw[fill] (360/\crater*\i: \diam) circle (2pt) +(360/\crater*\i: 0.4) node{$x^{\exp}$};
      }
    \end{tikzpicture}
    \caption{Undirected Schreier graph on $〈x〉\setminus\{1\}$ where $x^{13} = 1$,
			acted upon by $(ℤ/13ℤ)^*$, generated by $S=\{2,3,5\}$ (resp.
      blue, red and green edges).}
    \label{fig:cayley}
  \end{minipage}
\end{figure}

\section{Key exchange from isogeny graphs}
\label{sec:keyex}

We would like to instantiate the key exchange protocol of
Algorithm~\ref{alg:HHS-DH} with the PHS
$X = \Ell_q(\O)$ for the group $G = \Cl(\O)$, 
for some well chosen order $\O$ in a quadratic imaginary field. 
However, given a generic element of $\Cl(\O)$, 
the best algorithm~\cite{jao+soukharev10} to evaluate
its action on $\Ell_q(\O)$ has subexponential complexity in $q$,
making the protocol infeasible. 
The solution,
following Couveignes~\cite{cryptoeprint:2006:291},
is to fix a set $S$ of small prime ideals in $\O$,
whose action on $X$ can be computed efficiently,
and such that compositions of elements of $S$
cover the whole of $G$.
The action of an arbitrary element of $G$
is then the composition of a series of actions by small elements in $S$.
As Rostovtsev and Stolbunov~\cite{rostovtsev+stolbunov06} observed,
it is useful to visualise this decomposed action
as a walk in an isogeny graph.

\subsection{Walks in isogeny graphs}

Let $G$ be a group,
$X$ a PHS for $G$,
and~$S$ a subset of~$G$.
The Schreier graph $\Graph(G,S,X)$
is the labelled directed graph whose vertex set is~$X$, 
and where an edge labelled by $s∈S$
links $x_1$ to $x_2$ if and only if $s\cdot x_1 = x_2$.
It is isomorphic to a Cayley graph for $G$.
If $S$ is symmetric (that is, $S^{-1}=S$), 
then we associate the same label to $s$ and $s^{-1}$, 
making the graph undirected. 

A \emph{walk} in $\Graph(G,S,X)$ is a finite sequence
$(s_1,\ldots,s_n)$ of \emph{steps} in $S$. 
We define the action of this walk on $X$ as
\[
    (s_1,\ldots,s_n)·x 
    = 
    \big(\prod_{i=1}^n s_i\big)·x.
\]
In our application $G$ is abelian,
so the order of the steps $s_i$ does not matter.
We can use this action directly in the key exchange protocol
of Algorithm~\ref{alg:HHS-DH},
by simply taking private keys to be walks instead of elements in $G$.

\begin{example}
Figure~\ref{fig:cayley}
shows $\Graph(G,S,X)$ where $G=(ℤ/13ℤ)^*$, 
$S = \{2,3,5\}\cup\{2^{-1},3^{-1},5^{-1}\}$,
and $X = \langle{x}\rangle\setminus\{1\}$ 
is a cyclic group of order $13$, minus its identity element.
The action of $G$ on $X$ is exponentiation: $g·x=x^g$.
The action of $11$, which takes $x^k$ to $x^{11k}$,
can be expressed using the walks 
$(2,5,5)$,
or $(2^{-1},3^{-1})$,
or $(3,5)$,
for example.  Note that $5$ has order $4$ modulo
$13$, thus partitioning $〈x〉\setminus\{1\}$ into $3$ cycles of
length $4$.
\end{example}

Returning to the world of isogenies,
we now take
\begin{itemize}
    \item $X=\Ell_q(\O)$ as the vertex set, for some well-chosen $q$ and $\O$;
        in particular we require $\O$ to be the maximal order (see \S\ref{sec:sec}).
    \item $G=\Cl(\O)$ as the group acting on $X$;
    \item $S$ a set of ideals, whose norms are small Elkies primes in $\O$.
\end{itemize}
The graph $\Graph(G,S,X)$ is thus an isogeny graph, 
composed of many isogeny cycles (one for the norm of each prime in $S$) 
superimposed on the vertex set $\Ell_q(\O)$.
It is connected if $S$ generates $\Cl(\O)$.
Walks in $\Graph(G,S,X)$ are called \emph{isogeny walks}.

We compute the action of
an ideal $\frak s$ (a single \emph{isogeny step})
on an $x∈\Ell_q(\O)$ 
by choosing a representative curve $E$ with $x = j(E)$,
and computing an isogeny $ϕ_{\frak s}:E→E'$ from $E$
corresponding to $\frak{s}$;
the resulting vertex is $\frak s \cdot x = j(E')$.
The action of an isogeny walk $(\frak s_i)_i$
is then evaluated as the sequence of isogeny steps $ϕ_{\frak s_i}$. 
Algorithms for these operations are given in the next subsection. 

Using this ``smooth'' representation of elements
in $\Cl(\O)$ as isogeny walks
lets us avoid computing $\Cl(\O)$ and $\Ell_q(\O)$,
and avoid explicit ideal class arithmetic;
only isogenies between elliptic curves are computed.
In practice, we re-use the elliptic curve $E'$ from one step
as the $E$ in the next;
but we emphasize that
when isogeny walks are used for Diffie--Hellman,
the resulting public keys and shared secrets
are not the final elliptic curves,
but their $j$-invariants.

\subsection{Computing isogeny walks}

Since $\Cl(\O)$ is commutative,
we can break isogeny walks down into a succession of walks
corresponding to powers of single primes $\frak{s} = (\ell,\pi-\lambda)$;
that is, repeated applications of the isogenies $\phi_{\frak{s}}$.
Depending on $\frak{s}$,
we will compute each sequence of $\phi_{\frak s}$
using one of two different methods:
\begin{itemize}
    \item Algorithm~\ref{alg:ElkiesWalk} (\algstyle{ElkiesWalk})
        uses Algorithm~\ref{alg:ElkiesFirstStep}
        (\algstyle{ElkiesFirstStep})
        followed by a series of calls to Algorithm~\ref{alg:ElkiesNextStep} 
        (\algstyle{ElkiesNextStep}),
        both which use the \emph{modular polynomial} $\Phi_\ell(X, Y)$.
        This approach works for any $\frak s$.
    \item
        Algorithm~\ref{alg:VéluWalk} (\algstyle{VéluWalk})
        uses a series of calls to 
        Algorithm~\ref{alg:VéluStep} (\algstyle{VéluStep}).
        This approach, which uses torsion points on $E$, 
        can only be applied when
	    $\lambda$ satisfies certain properties.
\end{itemize}

Rostovtsev and Stolbunov only used analogues of 
Algorithms~\ref{alg:ElkiesFirstStep} and~\ref{alg:ElkiesNextStep}.
The introduction of \algstyle{VéluStep}, 
inspired by SIDH and related protocols
(and now a key ingredient in the CSIDH protocol~\cite{csidh}),
speeds up our protocol by a considerable factor;
this is the main practical contribution of our work. 

\begin{algorithm}
	\caption{\algstyle{ElkiesFirstStep}}
	\label{alg:ElkiesFirstStep}
	\Input{%
        $E\in\Ell_q(\O)$; 
        $(\ell,\lambda)$ encoding $\frak s = (\pi-\lambda,\ell)$
    }
	\Output{$j(\frak s\cdot E)$}
	$P\gets \Phi_\ell(X, j(E))$ 
    \;
    \label{line:roots}
    $\{j_1, j_2\} \gets \algstyle{Roots}(P,\F_q)$ 
    \;
    \label{line:kernel}
    $K\gets \algstyle{KernelPolynomial}(\algstyle{Isogeny}(E,j_1,\ell))$ 
    \tcp*{e.g.~BMSS algorithm~\cite{BMSS08}}
	$Q\gets$ a nonzero point in $K$ 
    \tcp*{e.g. $(x,y)\in E(\F_q[x,y]/(y^2-f_E(x),K(x)))$}
	\uIf{\label{line:direction} $\pi(Q) = [\lambda]Q$}{%
        \Return{$j_1$}
        \;
    }
	\Else{%
        \Return{$j_2$} \label{line:end_direction}
    }
\end{algorithm}

\begin{algorithm}
	\caption{\algstyle{ElkiesNextStep}}
	\label{alg:ElkiesNextStep}
    \Input{%
        $(\ell,\lambda)$ encoding $\frak{s} = (\pi-\lambda,\ell)$;
        $(j_0,j_1) = (j(E),j(\frak{s}\cdot E))$ for $E$ in $\Ell_q(\O)$
    }
    \Output{$j(\frak{s}\cdot\frak{s}\cdot E)$}
    \label{alg:ElkiesNextStep:step}
	$P\gets \Phi_\ell(X, j_1) / (X - j_0)$ 
    \;
	$j_2\gets \algstyle{Root}(P,\F_q)$ 
		\tcp*{It is unique}
	\Return{$j_2$}
    \;
\end{algorithm}

\begin{algorithm}
	\caption{\algstyle{ElkiesWalk}}
	\label{alg:ElkiesWalk}
	\Input{%
        $E\in \Ell_q(\O)$;
        $(\ell,\lambda)$ encoding $\frak s = (\pi - \lambda,\ell)$;
        $k\geq 1$
    }
	\Output{$\frak s^k \cdot E$}
	$j_0 \gets j(E)$ \;
    $j_1 \gets \algstyle{ElkiesFirstStep}(E, (\ell,\lambda))$ \;
	\For{$2\leq i \leq k$}{
        $(j_0, j_1) \gets (j_1, \algstyle{ElkiesNextStep}((\ell,\lambda),(j_0,j_1)))$ 
        \;
	}
    $E_R \gets \algstyle{EllipticCurveFromJInvariant}(j_1)$ 
    \;
    \label{line:twist}
    \If{\textbf{not} $\algstyle{CheckTrace}(E_R,t)$}{
        $E_R \gets \algstyle{QuadraticTwist}(E_R)$
        \;
    }
    \Return{$E_R$}
    \;
\end{algorithm}

\begin{algorithm}
	\caption{\algstyle{VéluStep}}
	\label{alg:VéluStep}
	\Input{%
        $E\in\Ell_q(\O)$; 
        $(\ell,\lambda)$ encoding $\frak s = (\pi-\lambda,\ell)$;
        $r > 0$;
        $C_r = \#E(\F_{q^r})$
    }
	\Output{$\frak s\cdot E$}
	\Repeat{$Q \neq 0_E$}{
        $P\gets \algstyle{Random}(E(\F_{q^r}))$ 
        \;
		$Q\gets [C_r/\ell] P$ \label{line:scalarmult}
        \;
	}
    $K\gets \prod_{i=0}^{(\ell-1)/2}(X - x([i]Q))$
    \tcp*{Kernel polynomial of isogeny} 
    $E_R\gets \algstyle{IsogenyFromKernel}(E,K)$ \label{line:quotient}
    \tcp*{Apply Vélu's formul\ae} 
	\Return{$E_R$}
    \;
\end{algorithm}

\begin{algorithm}
	\caption{\algstyle{VéluWalk}}
	\label{alg:VéluWalk}
	\Input{%
        $E\in \Ell_q(\O)$;
        $(\ell,\lambda)$ encoding $\frak s = (\ell,\pi - \lambda)$; 
        $k\geq 1$
    }
	\Output{$\frak s^k \cdot E$}
    $r\gets \algstyle{Order}(\lambda,\ell)$ 
    \tcp*{Precompute and store for each $(\ell,\lambda)$}
	$C_r \gets \# E(\F_{q^r})$ \label{line:card}
    \tcp*{Precompute and store for each $r$} 
    \For{$1 \le i \le k$}{
        $E \gets \algstyle{VéluStep}(E,(\ell,\lambda),r,C_r)$
        \;
    }
    \Return{$E$}
\end{algorithm}

\paragraph{Elkies steps.}
Algorithms~\ref{alg:ElkiesFirstStep}
and~\ref{alg:ElkiesNextStep} 
compute single steps in the $\ell$-isogeny graph.
Their correctness
follows from the definition of the modular polynomial $\Phi_\ell$:
a cyclic $\ell$-isogeny exists between two elliptic
curves $E$ and $E'$ if and only if $\Phi_\ell(j(E), j(E')) = 0$
(see~\cite[\S6]{schoof95} and~\cite[\S3]{Elkies98} for the relevant theory).
One may use the classical modular polynomials here,
or alternative, lower-degree modular polynomials
(Atkin polynomials, for example)
with minimal adaptation to the algorithms.
In practice, $\Phi_\ell$ is precomputed and stored:
several publicly available databases exist
(see~\cite{Echidna} and~\cite{SutherlandDatabase,BrokerLS12,BruinierOS16}, for example).

Given a $j$-invariant $j(E)$,
we can compute its two neighbours in the $\ell$-isogeny graph
by evaluating $P(X) = \Phi_\ell(j(E),X)$
(a polynomial of degree $\ell+1$),
and then computing its two roots in $\F_q$.
Using a Cantor--Zassenhaus-type algorithm,
this costs
$\softO(\ell\log q)$ $\F_q$-operations.

We need to make sure we step towards the neighbour in the correct direction.
If we have already made one such step, then this is easy:
it suffices to avoid backtracking.
Algorithm~\ref{alg:ElkiesNextStep} (\algstyle{ElkiesNextStep})
does this by
removing the factor corresponding to the previous $j$-invariant 
in Line~\ref{alg:ElkiesNextStep:step};
this algorithm can be used for all but the first of the steps
corresponding to $\frak{s}$.

It remains to choose the right direction in the first step 
for $\frak{s} = (\ell,\pi-\lambda)$.
In Algorithm~\ref{alg:ElkiesFirstStep}
we choose one of the two candidates for $\phi_{\frak s}$ arbitrarily,
and compute its kernel polynomial.
This costs $\softO(\ell)$ $\F_q$-operations
using the Bostan--Morain--Salvy--Schost algorithm~\cite{BMSS08}
with asymptotically fast polynomial arithmetic.
We then compute an element $Q$ of $\ker\phi_\frak{s}$
over an extension of $\F_q$ of degree at most $\frac{\ell-1}{2}$,
then evaluate $\pi(Q)$ and $[\lambda]Q$.
If they match, then we have chosen the right direction;
otherwise we take the other root of $P(X)$.

Algorithm~\ref{alg:ElkiesWalk} (\algstyle{ElkiesWalk}) 
combines these algorithms
to compute the iterated action of $\frak{s}$.
Line~\ref{line:twist} ensures that the curve returned
is the the correct component of the $\ell$-isogeny graph.
Both \algstyle{ElkiesFirstStep} and \algstyle{ElkiesNextStep}
cost $\softO(\ell\log q)$ $\F_q$-operations,
dominated by the calculation of the roots of $P(X)$.

\paragraph{Vélu steps.}

For some ideals $\frak{s} = (\ell,\pi-\lambda)$,
we can completely avoid modular polynomials,
and the costly computation of their roots,
by constructing $\ker\phi_{\frak{s}}$ directly from $\ell$-torsion points.
Let $r$ be the order of $\lambda$ modulo $\ell$;
then $\ker\phi_{\frak s} \subseteq E(\F_{q^r})$.
If $r$ is not a multiple of the order of the other eigenvalue $\mu$
of $\pi$ on $E[\ell]$,
then $E[\ell](\F_{q^r}) = \ker\phi_{\frak s}$.
Algorithm~\ref{alg:VéluStep} (\algstyle{VéluStep}) 
exploits this fact to construct a generator $Q$ of $\ker\phi_{\frak s}$
by computing a point of order $\ell$ in $E(\F_{q^r})$.
The roots of the kernel polynomial of $\phi_{\frak s}$
$x(Q), \ldots, x([(\ell-1)/2]Q)$.%
\footnote{If the order of $μ$ divides $r$,
  Algorithm~\ref{alg:VéluStep} can be extended as follows: take
  $P∈E[ℓ]$, and compute $π(P) - [μ]P$; the result is either zero, or
  an eigenvector for $λ$. %
  This is not necessary for any of the primes in our proposed
  parameters.}

Constructing a point $Q$ of order $\ell$ in $E(\F_{q^r})$
is straightforward:
we take random points and multiply by the cofactor $C_r/\ell$,
where $C_r := \#E(\F_{q^r})$.
Each trial succeeds with probability $1 - 1/\ell$.
Note that $C_r$ can be easily (pre)computed from the Frobenius trace $t$:
if we write $C_r = q - t_r + 1$ for 
$r > 0$ (so $t_1 = t$) and $t_0 = 2$,
then the $t_r$ satisfy the recurrence
$t_r = t\cdot t_{r-1} - q\cdot t_{r-2}$. 

We compute the quotient curve in Line~\ref{line:quotient} 
with Vélu's formul\ae~\cite{Velu71} in $O(\ell)$ $\F_q$-operations.
Since $\log C_r\simeq r\log q$, provided $\ell = O(\log q)$,
the costly step in Algorithm~\ref{alg:VéluStep} is the scalar
multiplication at Line~\ref{line:scalarmult}, which costs
$\softO(r^2\log q)$ $\F_q$-operations.

\paragraph{Comparing the costs.}
To summarize:
\begin{itemize}
\item Elkies steps cost $\softO(\ell\log q)$ $\F_q$-operations;
\item Vélu steps cost $\softO(r^2\log q)$ $\F_q$-operations,
    where $r$ is the order of $\lambda$ in $\Z/\ell\Z$.
\end{itemize}
In general $r = O(\ell)$, so Elkies steps should be preferred. 
However, when $r$ is particularly small
(and not a multiple of the order of the other eigenvalue),
a factor of $\ell$ can be saved using Vélu steps.
The value of $r$ directly depends on $\lambda$, which is in turn determined by
$\#E(\F_p)$ mod $\ell$. Thus, we see that better \algstyle{Step} performances
depend on the ability to find elliptic curves whose order
satisfies congruence conditions modulo small primes.
Unfortunately, we can only achieve this partially
(see \S\ref{sec:initcurve}), so the most efficient solution is
to use Vélu steps when we can, and Elkies steps for some other primes.

In practice, Algorithm~\ref{alg:VéluStep} can be improved
by using elliptic curve models with more efficient arithmetic.
In our implementation (see \S\ref{sec:exp}),
we used $x$-only arithmetic on Montgomery models~\cite{Montgomery87,CostelloSmith2017},
which also have convenient Vélu formul\ae~\cite{CostelloH17,Renes2018}.
Note that we can also avoid computing $y$-coordinates in
Algorithm~\ref{alg:ElkiesFirstStep}
at Line~\ref{line:direction} if $\lambda\neq\pm\mu$:
this is the typical case for Elkies steps, and we used this
optimization for all Elkies primes in our implementation.

\begin{remark}
    \label{rem:twist-trick}
    Note that, in principle, Algorithm~\ref{alg:VéluStep}, can only be
    used to walk in one direction $\frak s_λ=(ℓ,π-λ)$, and not in the
    opposite one $\frak s_μ=(ℓ,π-μ)$. Indeed we have assumed that
    $E[\frak s_λ]$ is in $E(\F_{q^r})$, while $E[\frak s_μ]$ is not.
    However, switching to a quadratic twist $\tilde{E}$ of $E$ over
    $\F_{q^r}$ changes the sign of the Frobenius eigenvalues, thus it
    may happen that $\tilde{E}[\frak s_{-μ}]$ is in
    $\tilde{E}(\F_{q^r})$, while $\tilde{E}[\frak s_{-λ}]$ is not. It
    is easy to force this behavior by asking that
    $p \equiv -1\pmod{\ell}$, indeed then $\lambda = -1/μ$.

    For these eigenvalue pairs we can thus walk in both directions
    using Vélu steps at no additional cost, following either the
    direction $λ$ on $E$, or the direction $-μ$ on a twist. In
    Algorithm~\ref{alg:VéluStep}, only the curve order and the random
    point sampling need to be modified when using quadratic twists.
\end{remark}

\subsection{Sampling isogeny walks for key exchange}

We now describe how keys are generated and exchanged in our protocol. 
Since the cost of the various isogeny walks depends on the ideals
chosen,
we will use adapted, or \emph{skewed}, 
smooth representations when sampling elements in $\Cl(\O)$
in order to minimize the total computational cost of a key exchange.

We take a (conjectural) generating set for $\Cl(\O)$ consisting of ideals over a set 
$S$ of small Elkies primes,
which we partition into three sets according to the step algorithms to be used.
We maintain three lists of tuples encoding these primes:

\begin{description}
    \item[$S_{VV}$]
        \label{case:velustep-sym}
        is a list of tuples $(\ell,\lambda,\mu)$
        such that the ideal $(\ell,\pi - \lambda)$ 
        and its inverse $(\ell,\pi-\mu)$
        are \emph{both} amenable to \algstyle{VéluStep}.
    \item[$S_{VE}$]
        \label{case:velustep-asym}
        is a list of tuples $(\ell,\lambda)$
        such that $(\ell,\pi-\lambda)$ is amenable to
        \algstyle{VéluStep} but its inverse $(\ell,\pi-\mu)$ is \emph{not}.
    \item[$S_{EE}$]
        \label{case:elkstep} 
        is a list of tuples $(\ell,\lambda,\mu)$
        such that \emph{neither} $(\ell,\pi - \lambda)$ nor $(\ell,\pi-\mu)$
        are amenable to \algstyle{VéluStep}.
\end{description}

In $S_{VV}$ and $S_{EE}$,
the labelling of eigenvalues as $\lambda$ and $\mu$ 
is fixed once and for all 
(that is, the tuples $(\ell,\lambda,\mu)$ and $(\ell,\mu,\lambda)$ do not
both appear).
This fixes directions in each of the $\ell$-isogeny cycles.
Looking back at Figure~\ref{fig:cycle}, 
for $\ell$ associated with $S_{EE}$ and $S_{VV}$, 
both directions in the $\ell$-isogeny graph will be available for use in
walks; for $S_{VE}$, only the Vélu direction will be used.

Each secret key in the cryptosystem is a walk in the isogeny graph.
Since the class group $\Cl(\O)$ is commutative, 
such a walk is determined by the multiplicities of the primes $\frak{s}$
that appear in it.
Algorithm~\ref{alg:isogeny-KeyGen} (\algstyle{KeyGen})
therefore encodes private-key walks as \emph{exponent vectors},
with one integer exponent for each tuple in $S_{VV}$, $S_{VE}$, and $S_{EE}$.
For a tuple $(\ell,\lambda,\mu)$,
\begin{itemize}
    \item
a positive exponent $k_\ell$ indicates a walk of $k_\ell$ $\ell$-isogeny
steps in direction $\lambda$;
\item
a negative exponent $-k_\ell$ indicates $k_\ell$
$\ell$-isogeny steps in direction $\mu$.
\end{itemize}
For the tuples $(\ell,\lambda)$ in $S_{VE}$, 
where we do not use the slower $\mu$-direction,
we only allow non-negative exponents.
We choose bounds $M_\ell$ on the absolute value of the exponents $k_\ell$
so as to minimize the total cost of computing isogeny walks, 
while maintaining a large keyspace.
As a rule, the bounds will be much bigger for the primes
in $S_{VV}$ and $S_{VE}$, where Vélu steps can be applied.

The public keys are $j$-invariants in $\F_q$,
so they can be stored in $\log_2 q$ bits;
the private keys are also quite compact,
but their precise size depends on the number of primes $\ell$
and the choice of exponent bounds $M_\ell$,
which is a problem we will return to in \S\ref{sec:exp}.

\begin{algorithm}
    \caption{%
        \algstyle{KeyGen} for cryptosystems in the isogeny graph on
        $\Ell_q(\O)$ with walks based on $S$, and initial curve $E_0$.
        The ideal lists $S_{EE}$, $S_{VV}$, and $S_{VE}$,
        and the walk bounds $M_\ell$, are system parameters.
    }
    \label{alg:isogeny-KeyGen}
    \Input{()}
    \Output{%
        A secret key $(k_\ell)_{\ell\in S}$ and the corresponding public
        key $j(E)$
    }
    $E \gets E_0$
    \;
    \For{$(\ell, \lambda, \mu)\in S_{EE}$}{%
        $k_\ell\gets \algstyle{Random}([-M_\ell, M_\ell])$ 
        \;
        \lIf{$k_\ell \ge 0$}{%
            $\nu \gets \lambda$
        }
        \lElse{%
            $\nu \gets \mu$
        }
        $E \gets \algstyle{ElkiesWalk}(E,(\ell,\nu),|k_\ell|)$
        \;
    }
    \For{$(\ell, \lambda, \nu) \in S_{VV}$}{%
        $k_\ell\gets \algstyle{Random}([-M_\ell, M_\ell])$ 
        \;
        \lIf{$k_\ell \ge 0$}{%
            $\nu \gets \lambda$
        }
        \lElse{%
            $\nu \gets \mu$
        }
        $E \gets \algstyle{VéluWalk}(E,(\ell,\nu),|k_\ell|)$
        \;
    }
    \For{$(\ell, \lambda)\in S_{VE}$}{%
        $k_\ell\gets \algstyle{Random}([0,M_\ell])$ 
        \;
        $E \gets \algstyle{VéluWalk}(E,(\ell,\lambda),k_\ell)$
        \;
    }
    \Return{$((k_\ell)_{\ell\in S},j(E))$}
\end{algorithm}

Algorithm~\ref{alg:isogeny-DH}
completes a Diffie--Hellman key exchange
by applying a combination of Elkies and Vélu walks
(Algorithms~\ref{alg:ElkiesWalk} and~\ref{alg:VéluWalk}, respectively).

\begin{algorithm}
    \caption{%
        \algstyle{DH} for the isogeny graph on $\Ell_q(\O)$
        with primes in $S$.
        The ideal lists $S_{EE}$, $S_{VV}$, and $S_{VE}$,
        and the walk bounds $M_\ell$, are system parameters.
        Public key validation is not included here,
        but (if desired) should be carried out 
        as detailed in \S\ref{sec:validation}.
    }
    \label{alg:isogeny-DH}
    \Input{%
        A private key \(k_A = (k_{A,\ell})_{\ell\in S}\)
        corresponding to a walk $(\frak{s}_1,\ldots,\frak{s}_n)$,
        and
        a public key $j_B = j(E_B)$ for $E_B \in \Ell_q(O)$
    }
    \Output{%
        A shared secret $j(\prod_{i=1}^n\frak{s_i}\cdot E_B)$
    }
    $E \gets \algstyle{EllipticCurveFromJInvariant}(j_B)$
    \;
    \If{\textbf{not} $\algstyle{CheckTrace}(E,t)$}{%
        $E \gets \algstyle{QuadraticTwist}(E)$
        \;
    }
    \For{$(\ell,\lambda,\mu) \in S_{EE}$}{
        \lIf{$k_{A,\ell} \ge 0$}{%
            $\nu \gets \lambda$
        }
        \lElse{%
            $\nu \gets \mu$
        }
        $E \gets \algstyle{ElkiesWalk}(E,(\ell,\nu),|k_{A,\ell}|)$
    }
    \For{$(\ell,\lambda,\mu) \in S_{VV}$}{
        \lIf{$k_{A,\ell} \ge 0$}{%
            $\nu \gets \lambda$
        }
        \lElse{%
            $\nu \gets \mu$
        }
        $E \gets \algstyle{VéluWalk}(E,(\ell,\nu),|k_{A,\ell}|)$
        \;
    }
    \For{$(\ell,\lambda) \in S_{VE}$}{
        $E \gets \algstyle{VéluWalk}(E,(\ell,\lambda),k_{A,\ell})$
        \;
    }
    \Return{$j(E)$}
\end{algorithm}

\section{Public parameter selection}
\label{sec:initcurve}

It is evident that the choice of public parameters
has a heavy impact on the execution time: 
smaller Elkies primes,
and smaller multiplicative orders of the Frobenius eigenvalues, 
will lead to better performance.
Since all of this information is contained in the value of $\# E(\F_q)$, 
we now face the problem of constructing ordinary elliptic curves 
of prescribed order modulo small primes. 
Unfortunately, and in contrast with the supersingular case,
no polynomial-time method to achieve this is known in general:
the CM method~\cite{AtkinMorain93,Sutherland2012},
which solves this problem
when the corresponding class groups are small,
is useless in our
setting (see \S\ref{sec:sec}).

In this section we describe how to use the Schoof--Elkies--Atkin (SEA) point counting
algorithm with early abort, combined with the use of certain modular curves,
to construct curves whose order satisfies some constraints
modulo small primes.
This is faster than choosing curves at random and computing their orders
completely until a convenient one is found, but it still does not allow us
to use the full power of Algorithm~\algstyle{VéluStep}.

\paragraph{Early-abort SEA.}
The SEA algorithm~\cite{schoof95,Morain95}
is the state-of-the-art point-count\-ing algorithm
for elliptic curves over large-characteristic finite fields.
In order to compute $N = \# E(\F_p)$, it computes $N$ modulo a series
of small Elkies primes~$\ell$, before combining the results via the CRT
to get the true value of $N$.

Cryptographers are usually interested in generating elliptic curves of
prime or nearly prime order, and thus without small prime factors.
While running SEA on random candidate curves,
one immediately detects if $N \equiv 0\pmod{\ell}$
for the small primes $\ell$;
if this happens then the SEA execution is aborted, and restarted with a new curve.

Here, the situation is the opposite: we \emph{want} elliptic curves
whose cardinality has many small prime divisors.
To fix ideas, we choose the 512-bit prime
\[
    p 
    := 
    7 \left(
        \prod_{2\leq\ell\leq 380,\ \ell \text{ prime}} \ell
    \right) - 1
    \,.
\]
Then, according to Remark~\ref{rem:twist-trick},
Algorithm~\algstyle{VéluStep} can be used for $\ell$-isogenies in both
directions for any prime $\ell\leq 380$, as soon as the order of its
eigenvalues is small enough.
We now proceed as follows:
\begin{itemize}
    \item 
        Choose a smoothness bound $B$ (we used $B = 13$).
    \item 
        Pick elliptic curves $E$ at random in $\F_p$, and use the SEA algorithm,
        aborting when any $\ell\leq B$ 
        with $\#E(\F_p) \not\equiv 0\pmod{\ell}$ is found.
    \item 
        For each $E$ which passed the tests above, 
        complete the SEA algorithm to compute $\#E(\F_p)$, 
        and estimate the key exchange running time 
        using this curve as a public parameter (see \S\ref{sec:exp}).
    \item
        The ``fastest'' curves now give promising candidates for $\#E(\F_p)$.
\end{itemize}

In considering the efficiency of this procedure, it is important to remark
that very few curves will pass the early-abort tests. 
The bound $B$ should be chosen 
to balance the overall cost of the first few tests with that of
the complete SEA algorithm for the curves which pass them. Therefore,
its value is somewhat implementation-dependent. 

\paragraph{Finding the maximal order.}
Once a ``good'' curve $E$ has been computed, we want to find a curve
$E_0$ having the same number of points, but whose endomorphism ring is maximal, and
to ensure that its discriminant is a large integer. Therefore, we attempt to
factor the discriminant $Δ_\pi$ of $\Z[\pi]$: if it is squarefree,
then $E$ already has maximal endomorphism ring, and in general the square
factors of $Δ_\pi$ indicate which ascending isogenies have to be computed
in order to find $E_0$.

\begin{remark}
    Factoring random 512-bit integers is not hard in general, 
    and discriminants of quadratic fields
    even tend to be slightly smoother than random integers.
    If a discriminant fails to be completely factored,
		a conservative strategy would be to discard it,
    but ultimately undetected large prime-square factors
    do not present a security issue
    because computing the possible corresponding large-degree isogenies 
    is intractable (see \S\ref{sec:sec}).
\end{remark}

\paragraph{Using the modular curve $X_1(N)$.}
Since we are looking for curves with smooth cardinalities, another
improvement to this procedure is available: instead of choosing elliptic
curves uniformly at random, we pick random candidates using
an equation for the modular curve $X_1(N)$~\cite{sutherland2012constructing},
which guarantees the existence of a rational $N$-torsion point
on the sampled elliptic curve.
This idea is used
in the procedure of selecting elliptic curves in the Elliptic Curve Method
for factoring~\cite{ECM20,GMP-ECM}.
In our implementation we used $N = 17$,
and also incorporated the existence test in~\cite{OKS00}
for Montgomery models for the resulting elliptic curves.

\paragraph{Results.}
We implemented this search using the Sage computer algebra system.
Our experiments were conducted on several machines running
Intel Xeon E5520 processors at 2.27GHz.
After 17,000 hours of CPU time, we found the Montgomery elliptic curve 
$
	E : y^2 = x^3 + A x^2 + x
$
over $\F_p$ with $p$ as above, and
\[
  A =\
  \begin{subarray}{l}
    108613385046492803838599501407729470077036464083728319343246605668887327977789 \\
    32142488253565145603672591944602210571423767689240032829444439469242521864171\,.
  \end{subarray}
\]
The trace of Frobenius $t$ of $E$ is
\[
    \scriptstyle -147189550172528104900422131912266898599387555512924231762107728432541952979290\,.
\]
There is a rational $\ell$-torsion point 
on $E$, or its quadratic twist, for each $\ell$ in
\[
  \{3, 5, 7, 11, 13, 17, 103, 523, 821, 947, 1723\}
  \,;
\]
each of these primes is Elkies.
Furthermore, $\End(E)$ is the maximal order, and its discriminant is
a 511-bit integer that has the following prime factorization:
\[
\begin{aligned}
    \scriptstyle -
    & \scriptstyle 2^3 \cdot 20507 \cdot 67429 \cdot 11718238170290677 \cdot 12248034502305872059 \\
    & \scriptstyle {}\cdot 60884358188204745129468762751254728712569\\
    & \scriptstyle {}\cdot 68495197685926430905162211241300486171895491480444062860794276603493\,.
\end{aligned}
\]
In \S\ref{sec:exp}, we discuss the practical performance
of our key-exchange protocol using these system parameters.
Other proposals for parameters are given in~\cite{memoire}.

\section{Security}
\label{sec:sec}

We now address the security of the CRS primitive, and derived
protocols. Intuitively, these systems rely on two assumptions:
\begin{enumerate}
    \item given two curves $E$ and $E'$ in $\Ell_q(\O)$, it is hard to
        find a (smooth degree) isogeny $ϕ:E→E'$; and
    \item the distribution on $\Ell_q(\O)$ induced by the random walks
        sampled in Algorithm~\ref{alg:isogeny-KeyGen} is computationally
        undistinguishable from the uniform distribution.
\end{enumerate}

We start by reviewing the known attacks for the first problem, both in
the classical and the quantum setting. Then, we formalize security
assumptions and give security proofs against passive adversaries.
Finally, we discuss key validation and protection against active
adversaries.

\subsection{Classical attacks}
\label{sec:classical-attacks}

We start by addressing the following, more general, problem:

\begin{problem}
\label{prob:isog}
  Given two ordinary elliptic curves $E,E'$ defined over a finite
  field $\F_q$, such that $\#E(\F_q)=\#E'(\F_q)$, find an isogeny walk
  $(ϕ_i)_{1≤i≤n}$ such that $ϕ_n∘\cdots∘ϕ_1(E)=E'$.
\end{problem}

The curves in Problem~\ref{prob:isog} are supposed to be sampled
uniformly, though this is never exactly the case in practice.
This problem was studied before the emergence of
isogeny-based cryptography~\cite{Gal,GHS,galbraith+stolbunov11},
because of its applications to conventional elliptic-curve
cryptography~\cite{GHS,teske06,jao+miller+venkatesan09}.
The algorithm with the best asymptotic complexity is due to Galbraith,
Hess and Smart~\cite{GHS}. It consists of three stages:
\begin{description}
\item[Stage 0.] Use walks of \emph{ascending} isogenies to reduce to the case where
  $\End(E)\cong\End(E')$ is the maximal order.
\item[Stage 1.] Start two random walks of horizontal isogenies 
  from $E$ and $E'$; detect the
  moment when they collide using a Pollard-rho type of algorithm.
\item[Stage 2.] Reduce the size of the obtained walk using
  index-calculus techniques.
\end{description}

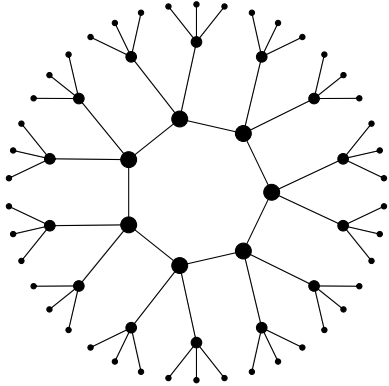
\begin{figure}
  \centering
  \begin{tikzpicture}
    \def\crater{7}
    \foreach \i in {1,...,\crater} {
      \draw[fill] (360/\crater*\i:1cm) circle (3pt);
      \draw (360/\crater*\i : 1cm) -- (360/\crater*\i+360/\crater : 1cm);
      \foreach \j in {-1,1} {
        \draw[fill] (360/\crater*\i : 1cm) -- (360/\crater*\i + \j*360/\crater/4 : 2cm) circle (2pt);
        \foreach \k in {-1,0,1} {
          \draw[fill] (360/\crater*\i + \j*360/\crater/4 : 2cm) --
          (360/\crater*\i + + \j*360/\crater/4 + \k*360/\crater/6 : 2.5cm) circle (1pt);
        }
      }
    }
  \end{tikzpicture}
  \caption{$3$-isogeny graph (\emph{volcano}) containing the curve
    with $j(E)=607$ over $\F_{6007}$. A larger vertex denotes a larger
    endomorphism ring.}
  \label{fig:volcano}
\end{figure}

To understand Stage~0, recall that all isogenous elliptic curves have
the same order, and thus the same trace $t$ of the Frobenius
endomorphism $π$. 
We know that
$\End(E)$ is contained in the ring of integers $\O_K$ of
$K=ℚ(\sqrt{Δ_π})$,
where $Δ_π=t^2-4q$ is the Frobenius discriminant.
As before we write $Δ_π=d^2Δ_K$, where $Δ_K$ is the
discriminant
of $\O_K$; then for any $ℓ\mid d$, the
$ℓ$-isogeny graph of~$E$ contains \emph{ascending} and
\emph{descending} $\ell$-isogenies; 
these graphs are referred to as 
\emph{volcanoes}~\cite{fouquet+morain02} (see Figure~\ref{fig:volcano}).
Ascending isogenies go from curves with smaller endomorphism rings to
curves with larger ones, and take us to a curve with $\End(E)≃\O_K$ in
$O(\log d)$ steps; they can be computed efficiently using the
algorithms
of~\cite{kohel,fouquet+morain02,ionica+joux13,defeo2016explicit}.
Assuming\footnote{%
    This is typical for isogeny-based protocols.
    No counter-example has ever been constructed.
}
all prime factors of $d$ are in $O(\log q)$,
we can therefore compute Stage~0
in time polynomial in $\log q$.

The set $\Ell_q(\O_K)$ has the smallest size among all sets
$\Ell_q(\O)$ for $\O⊂\O_K$, so it is always interesting to
reduce to it. This justifies using curves with maximal endomorphism
ring in the definition of the protocol in
\S\ref{sec:keyex}. When $Δ_π$ is square-free, $ℤ[π]$
is the maximal order, and the condition is automatically true.

The collision search in Stage~1 relies on the birthday paradox, and
has a complexity of $O(\sqrt{h(\O_K)})$.
It is known that, on average,
$h(\O_K)≈0.461\cdots\sqrt{|Δ_K|}$ (see~\cite[5.10]{Cohen1993}), and,
assuming the extended Riemann hypothesis, we even have a lower bound
(see~\cite{littlewood1928class})
\[h(\O_K) ≥ 0.147\cdots\frac{(1+o(1))\sqrt{|Δ_K|}}{\log\log|Δ_K|}.\]
Since $Δ_K\sim q$, we expect Stage~1 to take time $O(q^{1/4})$,
which justifies a choice of $q$ four times as large as the
security parameter.  Unfortunately, class numbers are notoriously
difficult to compute, the current record being
for a discriminant of 300 bits~\cite{10.1007/978-3-642-14081-5_15}.
Computing class numbers for ${\sim 500}$-bit discriminants seems to be expensive,
albeit feasible; thus, we can only rely on these heuristic arguments
to justify the security of our proposed parameters.

The horizontal isogeny produced by Stage~1 is represented by an ideal
constructed as a product of exponentially many small ideals.
Stage~2 converts this into a sequence of small ideals of length
polynomial in \(\log q\).
Its complexity is bounded by that of Stage~1,
so it has no impact on our security estimates.

\begin{remark}
  The Cohen--Lenstra heuristic~\cite{10.1007/BFb0099440} predicts that
  the odd part of $\Cl(\O_K)$ is cyclic with overwhelming
  probability, and other heuristics~\cite{10.1007/3-540-44448-3_18}
  indicate that $h(\O_K)$ is likely to have a large prime factor.
  However, since there is no known way in which the group structure of
  $\Cl(\O_K)$ can affect the security of our protocol, we can 
  disregard this matter. No link between the group structure
	of $E(\F_q)$ itself and the security is known, either.
\end{remark}

\subsection{Quantum attacks}
\label{sec:quantum-attacks}

On a quantum computer, an attack with better asymptotic complexity is
given by Childs, Jao and Soukharev
in~\cite{childs2014constructing}. It consists of two algorithms:
\begin{enumerate}
\item A (classical) algorithm that takes as input an elliptic curve
  $E∈\Ell_q(\O)$ and an ideal $\frak a∈\Cl(\O)$, and outputs the curve
  $\frak a·E$;
\item A generic quantum algorithm for the dihedral hidden subgroup
  problem (dHSP), based upon previous work of Kuperberg~\cite{Kup,Kuperberg2013} and
  Regev~\cite{regev04}.
\end{enumerate}

The ideal evaluation algorithm has sub-exponential complexity
$L_q(\frac{1}{2},\frac{\sqrt{3}}{2})$. However, after a subexponential-time
classical precomputation, any adversary can know the exact class group structure;
in that case, this ideal evaluation step could possibly be performed in polynomial
time (and non-negligible success probability)
using LLL-based methods, as discussed in~\cite{Stolbunov2012}
and~\cite[\S5]{cryptoeprint:2006:291}.

The dHSP algorithm uses the
ideal evaluation algorithm as a (quantum) black box, the number of
queries depending on the variant. Childs--Jao--Soukharev gave two
versions of this algorithm, Kuperberg's~\cite{Kup} and Regev's\cite{regev04}.
However, both are superseded by Kuperberg's recent work~\cite{Kuperberg2013}:
his new algorithm solves the dHSP in any abelian group of order $N$
using $2^{O(\sqrt{\log N})}$ quantum queries and classical space,
but only $O(\log N)$ quantum space. 
Given this estimate, we expect the bit size of $q$ to grow at
worst like the square of the security parameter.

Unfortunately, the analysis of Kuperberg's new algorithm is only asymptotic, 
and limited to $N$ of a special form; it cannot be directly used
to draw conclusions on concrete cryptographic parameters at this stage,
especially since the value of the constant hidden by the $O()$ in the
exponent is unclear.
Thus, it is hard to estimate the impact of this attack
at concrete security levels
such as those required by NIST~\cite{NIST2016}.

Nevertheless, we remark that the first version of Kuperberg's
algorithm, as described in~\cite[Algorithm~5.1 and
Remark~5.2]{regev04} requires $O(2^{3\sqrt{\log N}}\log N)$ black-box
queries and $\sim 2^{3\sqrt{\log N}}$ qubits of memory.  Although the
quantum memory requirements of this algorithm are rather high, we will
take its query complexity as a crude lower bound for the complexity of
Kuperberg's newer algorithm in the general case.  Of course, this
assumption is only heuristic, and should be validated by further study
of quantum dHSP solvers; at present time, unfortunately, no precise
statement can be made.

Table~\ref{tab:sizes} thus proposes various parameter sizes, with
associated numbers of quantum queries based on the observations above;
we also indicate the estimated time to (classically) precompute the
class group structure according
to~\cite{10.1007/978-3-642-14081-5_15}.\footnote{%
  Computing the class group structure is an instance of
  the hidden subgroup problem, and thus can be solved in quantum
  polynomial time by Shor's algorithm.
}
Whenever the quantum query
complexity alone is enough to put a parameter in one of NIST's
security categories~\cite{NIST2016}, we indicate it in the table. We
believe that using query complexity alone is a very conservative
choice, and should give more than enough confidence in the
post-quantum security properties of our scheme.

The system parameters we proposed in \S\ref{sec:initcurve}
correspond to the first line of Table~\ref{tab:sizes}, 
thus offering at least 56-bit quantum 
and 128-bit classical security.

\begin{table}
    \renewcommand{\arraystretch}{1.4}
    \centering
    \begin{tabular}{c@{\;}|@{\;}c@{\;}|@{\;}c@{\;}|@{\;}c@{\;}|@{\;}c@{\;}|@{\;}c}
        $\log Δ_K$ & $\log h(\O_K)$
        & \parbox{10ex}{\centering classical security}
        & $L_{|Δ_K|}(1/2,1)$ 
        & \parbox{10ex}{\centering quantum queries}
        & \parbox{10ex}{\centering NIST category} \\
        \hline
        $512$  &  $256$ & $2^{128}$ &  $2^{56.6}$ & $> 2^{56}$ \\
        $688$  &  $344$ & $2^{172}$ &  $2^{67.0}$ & $> 2^{64}$ & 1\\
        $768$  &  $384$ & $2^{192}$ &  $2^{71.4}$ & $> 2^{67}$ & 1\\
        $1024$ &  $512$ & $2^{256}$ &  $2^{84.2}$ & $> 2^{76}$ & 1\\
        $1656$ &  $828$ & $2^{414}$ & $2^{110.8}$ & $> 2^{96}$ & 3\\
        $3068$ & $1534$ & $2^{767}$ & $2^{156.9}$ & $> 2^{128}$ & 5
        \\
        \hline
    \end{tabular}
    \smallskip
    \caption{Suggested parameter sizes and associated classical
      security, class group computation time, and query complexity,
      using the heuristic estimations of
      \S\ref{sec:quantum-attacks}.}
    \label{tab:sizes}
\end{table}

\subsection{Security proofs}
\label{sec:proofs}

We now formalize the assumptions needed to prove the security of the
key exchange protocol, and other derived protocols such as PKEs and
KEMs, in various models. Given the similarity with the classical
Diffie--Hellman protocol on a cyclic group, our assumptions are
mostly modeled on those used in that context. Here we are
essentially following the lead of
Couveignes~\cite{cryptoeprint:2006:291} and
Stolbunov~\cite{Stol,Stolbunov2012}.
However, we take their analyses a
step further by explicitly modeling the hardness of distinguishing
random walks on Cayley graphs from the uniform distribution: this
yields stronger proofs and a better separation of security concerns.

For the rest of this section $q$ is a prime power, $\O$ is an order in
a quadratic imaginary field with discriminant $Δ\sim q$, $\Cl(\O)$ is
the class group of $\O$, $\Ell_q(\O)$ is the (non-empty) set of
elliptic curves with complex multiplication by $\O$, and $E_0$ is a
fixed curve in $\Ell_q(\O)$. Finally, $S$ is a set of ideals of $\O$
with norm polynomial in $\log q$, and $σ$ is a probability
distribution on the set $S^*$ of isogeny walks (i.e. finite
sequences of elements in $S$) used to sample secrets in the key exchange
protocol.  
We write $x\rand{σ} X$ for an element taken from a set $X$
according to $σ$, 
and 
$x\uni X$ for an element taken according to the uniform distribution.

Our security proofs use four distributions on $\Ell_q(\O)^3$:
\begin{align*}
    \mathcal{G}_{q,Δ} 
    & := 
    \left\{
        (\frak a·E_0,\frak b·E_0,\frak{ab}·E_0)
        \suchthat 
        \frak a,\frak b\uni\Cl(\O)
    \right\}
    \,, \\
    \mathcal{W}_{q,Δ,σ} 
    & := 
    \left\{
        \bigl((\frak a_i)_i·E_0,(\frak b_j)_j·E_0,(\frak a_i)_i·(\frak b_j)_j·E_0\bigr)
        \suchthat 
        (\frak a_i)_i,(\frak b_j)_j\rand{σ}S^*
    \right\}
    \,, \\
    \mathcal{R}_{q,Δ,σ} 
    & := 
    \left\{
        \bigl((\frak a_i)_i·E_0,(\frak b_i)_i·E_0,E'\bigr)
        \suchthat
        (\frak a_i)_i,(\frak b_i)_i\rand{σ}S^*,\; E'\uni\Ell_q(\O)
    \right\}
    \,, \\
    \mathcal{U}_{q,Δ} 
    & :=
    \left\{
        (E_a,E_b,E_{ab}) 
        \suchthat 
        E_a,E_b,E_{ab}\uni\Ell_q(\O)
    \right\}
    \,.
\end{align*}

The assumption needed to prove security of the protocols is 
the hardness of a problem analogous to 
the classic Decisional Diffie--Hellman (DDH) problem.

\begin{definition}[Isogeny Walk DDH (IW-DDH)]
    Given a triplet of curves $(E_a,E_b,E_{ab})$
    sampled with probability $\frac{1}{2}$ 
    from $\mathcal{R}_{q,\Delta,\sigma}$
    and $\frac{1}{2}$ from $\mathcal{W}_{q,\Delta,\sigma}$,
    decide from which it was sampled. 
\end{definition}

We split this problem into two finer-grained problems. 
The first is that of distinguishing between 
commutative squares sampled uniformly at random 
and commutative squares sampled from the distribution $σ$.

\begin{definition}[Isogeny Walk Distinguishing (IWD)]
    Given a triplet of curves $(E_a,E_b,E_{ab})$
    sampled with probability 
    $\frac{1}{2}$ from $\mathcal{W}_{q,Δ,σ}$
    and 
    $\frac{1}{2}$ from $\mathcal{G}_{q,Δ}$,
    decide from which it was sampled.
\end{definition}

The second problem is a group-action analogue of DDH.
It also appears in~\cite{cryptoeprint:2006:291} under the
name \emph{vectorization}, and in~\cite{Stol,Stolbunov2012} under the
name DDHAP.
 
\begin{definition}[Class Group Action DDH (CGA-DDH)]
    Given a triplet of curves $(E_a,E_b,E_{ab})$
    sampled with probability 
    $\frac{1}{2}$ from $\mathcal{G}_{q,Δ}$
    and 
    $\frac{1}{2}$ from $\mathcal{U}_{q,Δ}$,
    decide from which it was sampled.
\end{definition}

We want to prove the security of protocols based on the primitive of
\S\ref{sec:keyex} under the CGA-DDH and IWD assumptions
combined. To do this we give a lemma showing that CGA-DDH and IWD
together imply IW-DDH. The technique is straightforward: we use an
IW-DDH oracle to solve both the CGA-DDH and IWD problems, showing that
at least one of the two must be solvable with non-negligible
advantage. The only technical difficulty is that we need
an efficient way to simulate the uniform distribution on $\Ell_q(\O)$;
for this, we use another Cayley graph on $\Ell_q(\O)$,
with a potentially larger edge set, that is proven
in~\cite{jao+miller+venkatesan09} to be an expander 
under the generalized Riemann hypothesis (GRH).

We let $\Adv[A]{IW-DDH}$ be the \emph{advantage} of an adversary $A$
against IW-DDH, defined as the probability that $A$ answers correctly,
minus $1/2$:
\[
    2\Adv[A]{IW-DDH} 
    = 
    \Proba\bigl[A(\mathcal{R}_{q,Δ,σ}) = 1\bigr] 
    -
    \Proba\bigl[A(\mathcal{W}_{q,Δ,σ}) = 1\bigr]
    \,.
\]
We define $\Adv[A]{CGA-DDH}$ and $\Adv[A]{IWD}$ similarly. 
Switching answers if needed, we can assume all advantages are positive.
We let $\Adv{X}(t)$ denote the maximum of $\Adv[A]{X}$ over all
adversaries using at most $t$ resources (running time, queries, etc.).

\begin{lemma}
\label{lem:adv}
    Assuming GRH, for $q$ large enough and for any bound $t$ on running
    time, and for any $\epsilon>0$,
    \[
        \Adv{IW-DDH}(t) 
        ≤ 
        2\Adv{IWD}(t+\poly(\log q, \log\epsilon)) 
        + 
        \Adv{CGA-DDH}(t) 
        + 
        \epsilon
        \,.
    \]
\end{lemma}
\begin{proof}[Sketch]
    We start with an adversary $A$ for IW-DDH, and we construct two simulators
    $S$ and $T$ for CGA-DDH and IWD respectively.
    \begin{itemize}
        \item
            The simulator $S$ simply passes its inputs to $A$,
            and returns $A$'s response.
        \item
            The simulator $T$ receives a triplet $(E_a,E_b,E_{ab})$ taken from
            $\mathcal{G}_{q,Δ}$ or $\mathcal{W}_{q,Δ,σ}$, and flips a coin
            to decide which of the two following actions it will do:
            \begin{itemize}
                \item forward $(E_a,E_b,E_{ab})$ to $A$, and return the bit
                    given by $A$; or
                \item generate a random curve $E_c∈\Ell_q(\O)$, forward
                    $(E_a,E_b,E_c)$ to $A$, and return the opposite bit to the one
                    given by $A$.
            \end{itemize}
    \end{itemize}
  
    The curve $E_c$ must be sampled from a distribution close to uniform
    for the simulator $T$ to work. The only way at our disposal to sample
    $E_c$ uniformly would be to sample a uniform $\frak c∈\Cl(\O)$ and take
    $E_c=\frak c·E_0$, but this would be too costly. Instead we
    use~\cite[Theorem~1.5]{jao+miller+venkatesan09},
	combined with standard results about random walks in
	expander graphs (for instance, an easy adaptation of the proof
	of~\cite[Lemma~2.1]{jao+miller+venkatesan09}), to sample $E_c$ so
    that any curve in $\Ell_q(\O)$ is taken with probability between
    $(1-\epsilon)/h(\O)$ and $(1+\epsilon)/h(\O)$,
	using only $\poly(\log q, \log\epsilon)$ operations.
	We can consider this sampling as follows:
	with probability $1-\epsilon$, sample $E_c$ uniformly,
	and with probability $\epsilon$ sample it from
	an unknown distribution.

    Now, if $T$ forwarded $(E_a,E_b,E_{ab})$ untouched, then we immediately get
    \begin{align*}
        2\Adv[T]{IWD} 
        & = 
        \Adv[A]{IW-DDH} - \Adv[S]{CGA-DDH}
        \,;
        \intertext{if $T$ forwarded $(E_a,E_b,E_c)$, then we get}
        2\Adv[T]{IWD} 
        & \geq 
        \Adv[A]{IW-DDH} - (1-\epsilon)\Adv[S]{CGA-DDH} - \epsilon
        \,.
    \end{align*}
    Averaging over the two outcomes concludes the proof.
    \qed
\end{proof}

Finally, we define an isogeny-walk analogue of the classic 
Computational Diffie--Hellman (CDH) problem for groups.
Using the same techniques as above, we can prove the
security of the relevant protocols based only on CGA-CDH and IWD,
without the generalized Riemann hypothesis.

\begin{definition}[Class Group Action CDH (CGA-CDH)]
    Given $E_a=\frak a·E_0$ and $E_b=\frak b·E_0$ 
    with $\frak a,\frak b\uni\Cl(\O)$, 
    compute the curve $E_{ab}=\frak{ab}·E_0$.
\end{definition}

Stolbunov proved
the security of HHS Diffie--Hellman under the equivalent of CGA-DDH~\cite{Stol}.
Repeating the same steps, we can prove the following theorem.

\begin{theorem}
    If the CGA-DDH and IWD assumptions hold, assuming GRH, 
    the key-agreement protocol defined by 
    Algorithms~\ref{alg:isogeny-KeyGen} and~\ref{alg:isogeny-DH}
    is session-key secure in the authenticated-links adversarial model 
    of Canetti and Krawczyk~\cite{canetti}.
\end{theorem}

Similarly, we can prove the IND-CPA security of the hashed ElGamal
protocol derived from Algorithm~\ref{alg:isogeny-KeyGen} by replicating the
techniques of e.g.~\cite[\S20.4.11]{galbraith2012mathematics}.

\begin{theorem}
    Assuming CGA-CDH and IWD, the hashed ElGamal protocol derived from
    Algorithms~\ref{alg:isogeny-KeyGen} and~\ref{alg:isogeny-DH}
    is IND-CPA secure in the random oracle model.
\end{theorem}

\paragraph{A heuristic discussion of the IWD assumption.}

From its very definition, the IWD problem depends on
the probability distribution $σ$ we use to sample
random walks in the isogeny graph. In this paragraph,
we provide heuristic arguments suggesting that 
the IWD instances generated by Algorithm~\ref{alg:isogeny-DH} are hard,
provided
\begin{enumerate}
    \item the keyspace size is at least $\sqrt{|Δ_K|}$, and
    \item $S$ is \emph{not too small}, i.e. the number of
        isogeny degrees used is in $\Omega(\log q)$.
\end{enumerate}

Proving rapid mixing of isogeny walks with such
parameters seems out of reach at present, even under number-theoretic
hypotheses such as GRH. The best results available,
like~\cite[Theorem~1.5]{jao+miller+venkatesan09}
(used in the proof of Lemma~\ref{lem:adv}),
typically require isogeny degrees in $\Omega((\log q)^B)$
for some $B>2$, and fully random walks that are not,
for example, skewed towards smaller-degree isogenies.

However, numerical evidence suggests that
these theoretical results are too weak. In~\cite[7.2]{jao+miller+venkatesan09},
it is asked whether an analogue of the previous theorem would
be true with the sole constraint $B>1$. In~\cite[Section~3]{GHS},
it is mentioned that many fewer split primes are needed to walk
in the isogeny graph than theoretically expected.
Practical evidence also suggests that the rapid mixing
properties are not lost with skewed random walks:
such walks are used in~\cite{galbraith+stolbunov11}
to accelerate an algorithm solving Problem~\ref{prob:isog}.
We believe that these experiments can bring some evidence
in favor of relying on the IWD assumptions with more aggressive
parameters than those provided by GRH, although further investigation
is required.

\subsection{Key validation and active security}
\label{sec:validation}

Modern practice in cryptography mandates the use of stronger security
notions than IND-CPA.  From the DLP assumption, it is easy to
construct protocols with strong security against active adversaries.
For example, it is well-known that the hashed ElGamal KEM achieves
IND-CCA security in the random oracle model under various
assumptions~\cite{10.1007/3-540-45353-9_12,cryptoeprint:1999:007,doi:10.1137/S0097539702403773}.

All of these constructions crucially rely on \emph{key validation}:
that is, Alice must verify that the public data sent by Bob defines valid
protocol data (e.g., valid elements of a cyclic group), or abort if
this is not the case.
Failure to perform key validation may result in catastrophic attacks,
such as small subgroup~\cite{10.1007/BFb0052240}, invalid
point~\cite{10.1007/3-540-44598-6_8}, and invalid curve
attacks~\cite{Ciet2005}.  

In our context, key validation amounts to
verifying that the curve sent by Bob really is an element of
$\Ell_q(\O_K)$. Failure to do so exposes Alice to an \emph{invalid
  graph attack}, where Bob forces Alice onto an isogeny class with much
smaller discriminant, or different Elkies primes, and learns something
on Alice's secret.

Fortunately, key validation is relatively easy for protocols based on
the CRS primitive. 
All we need to check is that the received $j$-invariant
corresponds to a curve with the right order, and with maximal endomorphism ring. 

\paragraph{Verifying the curve order.} Since we already know the trace $t$
of the Frobenius endomorphism of all curves in $\Ell_q(\O)$, 
we only need to check that the given $E$ has order $q+1-t$.
Assuming that $E$ is cyclic, or contains a
cyclic group of order larger than $4\sqrt{q}$, a very efficient
randomized algorithm consists in taking a random point $P$ and
verifying that it has the expected order.  This task is easy if the
factorization of $q+1-t$ is known.

Concretely, the curve given in \S\ref{sec:initcurve} has order
\[
    N 
    = 
    2^2 · 3^2 · 5 · 7 · 11 · 13^2 · 17 · 103 · 523 · 821 · 1174286389
    · (\text{432-bit prime})
    \,,
\] %
and its group structure is
$ℤ/2ℤ×ℤ/\frac{N}{2}ℤ$. To check that a curve is in the same isogeny
class, we repeatedly take random points until we find one of order~$N/2$.

\paragraph{Verifying the endomorphism ring level.} 
The curve order verification proves that $\End(E)$ 
is contained between $ℤ[π]$ and $\O_K$. 
We have already seen that there is only a finite number of
possible rings: their indices in $\O_K$ must divide $d$ where $d^2=Δ_π/Δ_K$.
Ascending and descending isogenies connect curves with different endomorphism
rings, thus we are left with the problem of verifying that $E$ is on
the crater of any $ℓ$-volcano for $ℓ\mid d$.
Assuming no large prime divides $d$, this check can be
accomplished efficiently by performing random walks in the volcanoes,
as described in~\cite[\S4.2]{kohel} or~\cite{fouquet+morain02}. 
Note that if we choose $Δ_π$ square-free, 
then the only possible endomorphism ring is $\O_K$,
and there is nothing to be done.

Concretely, for the curve of \S\ref{sec:initcurve}
we have $Δ_π/Δ_K=2^2$,
so there are exactly two possible endomorphism rings. Looking at
the action of the Frobenius endomorphism, we see that $\End(E)=\O_K$
if and only if $E[2]≃(ℤ/2ℤ)^2$.

\begin{example}
    Let $p$ and $\O$ be as in \S\ref{sec:initcurve}.
    Suppose we are given the value
    \[
        \alpha = 
        \begin{subarray}{l}
            67746537624003763704733620725115945552778190049699052959500793811735672493775
            \\
            18737748913882816398715695086623890791069381771311397884649111333755665289025
        \end{subarray}
    \]
    in $\F_p$.  It is claimed that $\alpha$ is in $\Ell_p(\O)$;
    that is, it is a valid public key for the system with parameters
    defined in \S\ref{sec:initcurve}.
    Following the discussion above,
    to validate $\alpha$ as a public key,
    it suffices to exhibit a curve with $j$-invariant $\alpha$,
    full rational $2$-torsion,
    and a point of order $N/2$.
    Using standard formul\ae{},
    we find that the two $\F_p$-isomorphism classes of elliptic curves
    with $j$-invariant $\alpha$
    are represented by the Montgomery curve
    $E_\alpha/\F_p: y^2 = x(x^2 + Ax + 1)$
    with
    \[
        A = 
        \begin{subarray}{l}
            41938099794353656685283683753335350833889799939411549418804218343694887415884
            \\
            66125999279694898695485836446054238175461312078403116671641017301728201394907
        \end{subarray}
    \]
    and its quadratic twist $E_\alpha'$.
    Checking the $2$-torsion first,
    we have $E_\alpha[2](\F_p) \cong E_\alpha'[2](\F_p) \cong (\Z/2\Z)^2$,
    because $A^2 - 4$ is a square in $\F_p$.
    Trying points on $E_\alpha$,
    we find that $(23,\sqrt{23(23^2 + 23A + 1)})$ in $E_\alpha(\F_p)$
    has exact order $N/2$.
    We conclude that $\End(E_\alpha) = \O$,
    so $\alpha$ is a valid public key.
    (In fact, $E_\alpha$ is connected to the initial curve
    by a single $3$-isogeny step.)
\end{example}

\paragraph{Consequences for cryptographic constructions.}
Since both of the checks above can be done much more efficiently 
than evaluating a single isogeny walk, 
we conclude that key validation is not only possible,
but highly efficient for protocols based on the CRS construction. 
This stands in stark contrast to the case of SIDH,
where key validation is known to be
problematic~\cite{galbraithsecurity}, and even conjectured to be as
hard as breaking the system~\cite{cryptoeprint:2018:336}.

Thanks to this efficient key validation, 
we can obtain CCA-secure encryption from the
CRS action without resorting to generic transforms such as
Fujisaki--Okamoto~\cite{10.1007/3-540-48405-1_34}, unlike the case of
SIKE~\cite{SIKE,10.1007/978-3-319-70500-2_12}. This in turn enables
applications such as non-interactive key exchange, for which no
practical post-quantum scheme was known prior to~\cite{csidh}.

\section{Experimental results}
\label{sec:exp}

In order to demonstrate that our protocol is usable at
standard security levels, we implemented it in the Julia programming
language. This proof of concept also allowed us to estimate 
isogeny step costs, which we needed to generate the
initial curve in \S\ref{sec:initcurve}.  We developed several
Julia packages\footnote{The main code is available at
  \url{https://github.com/defeo/hhs-keyex/}, and the additional
  dependencies at \url{https://github.com/defeo/EllipticCurves.jl/}
  and \url{https://github.com/defeo/ClassPolynomials.jl/}.}, built
upon the computer algebra package Nemo~\cite{nemo}.
Experiments were conducted
using Julia 0.6 and Nemo 0.7.3 on Linux, with an Intel Core i7-5600U cpu at 2.60GHz.

Consider the
time to compute one step for an ideal $\frak s = (\ell,\pi-\lambda)$.
Using Elkies steps, this is approximately
the cost of finding the roots of the modular polynomial:
roughly $0.017\cdot\ell$ seconds in our implementation.
Using Vélu steps, the cost is approximately
that of one scalar multiplication in $E(\F_{q^r})$;
timings for the extension degrees \(r\) relevant to our parameters
appear in Table~\ref{tab:time-vs-r}.

\begin{table}
    \centering
    \begin{tabular}{r@{\;}|@{\;}c@{\ \ }c@{\ \ }c@{\ \ }c@{\ \ }c@{\ \ }c@{\ \ }c}
        $r$ & 1 & 3 & 4 & 5 & 7 & 8 & 9 \\
        \hline
        time (s) & 0.02 & 0.10 & 0.15 & 0.24 & 0.8 & 1.15 & 1.3
    \end{tabular}
    \smallskip
    \caption{Timings for computing scalar multiplications in
    $E(\F_{p^r})$, the dominant operation in \algstyle{VéluStep}
    (Algorithm~\ref{alg:VéluStep}), as a function of the extension
    degree $r$.}
    \label{tab:time-vs-r}
\end{table}

Using this data, finding efficient walk length bounds $M_\ell$
offering a sufficient keyspace size is easily seen to be an
integer optimization problem. 
We used the following heuristic procedure
to find a satisfactory solution.
Given a time bound $T$, let \algstyle{KeySpaceSize}$(T)$ be
the keyspace size obtained when each $M_\ell$ is the greatest such
that the \emph{total time spent on $\ell$-isogenies} is less than $T$.
Then, if $n$ is the (classical) security parameter,
we look for the \emph{least $T$} such that
$
	\algstyle{KeySpaceSize}(T)\geq 2^{2n}
$
(according to \S\ref{sec:sec}), using binary search.
While the $M_\ell$
we obtain are most likely not the best possible, intuitively
the outcome is not too far from optimal.

In this way, we obtain a proposal for the walk length bounds $M_\ell$
to be used in Algorithm~\ref{alg:isogeny-KeyGen} along with the curve
found in \S\ref{sec:initcurve}, to achieve 128-bit
classical security. Table~\ref{tab:VéluSteps}
lists the isogeny degrees amenable to Algorithm~\ref{alg:VéluStep},
each with the corresponding extension degree $r$ 
(a star denotes that the twisted curve
allows us to use both directions in the isogeny graph,
as in Remark~\ref{rem:twist-trick}).
Table~\ref{tab:ElkiesSteps}
lists other primes for which we apply Algorithm~\ref{alg:ElkiesWalk}.

\begin{table}
    \centering
    \begin{tabular}{l|@{\;}r@{\;}|@{\;}l@{\;}||l|@{\;}r@{\;}|@{\;}l@{\;}||@{\;}l|r@{\;}|@{\;}l}
        $r$ & $M_\ell$ & $\ell$
        &
        $r$ & $M_\ell$ & $\ell$
        &
        $r$ & $M_\ell$ & $\ell$
        \\
        \hline
        1* & 409 & 3, 5, 7, 11, 13, 17, 103
        &
        4 & 54 & 1013, 1181
        &
        8 & 7 & 881
        \\
        1 & 409 & 523, 821, 947, 1723
        &
        5 & 34 & 31*, 61*, 1321
        &
        9 & 6 & 37*, 1693
        \\
        3 & 81 & 19*, 661
        &
        7 & 10 & 29*, 71*, 547
        & & 
        \\
        \hline
    \end{tabular}
    \smallskip
    \caption{Primes $\ell$ amenable to Algorithm~\ref{alg:VéluStep}
		(\algstyle{VéluStep}) for our candidate isogeny graph,
    with corresponding extension degrees $r$ 
    and proposed walk length bounds $M_\ell$.}
    \label{tab:VéluSteps}
\end{table}

\begin{table}
    \centering
    \begin{tabular}{r@{\;}|@{\;}l@{\;}||@{\;}r@{\;}|@{\;}l@{\;}||@{\;}r@{\;}|@{\;}l@{\;}}
        $M_\ell$ & $\ell$
        &
        $M_\ell$ & $\ell$
        &
        $M_\ell$ & $\ell$
        \\
        \hline
        20 & 23
        &
        6  & 73
        &
        2  & 157, 163, 167, 191, 193, 197, 223, 229
        \\
        11 & 41
        &
        5  & 89
        &
        1  & 241, 251, 257, 277, 283, 293, 307
        \\
        10 & 43
        &
        4  & 107, 109, 113
        &
        1  & 317, 349, 359
        \\
        9 & 47
        &
        3  & 131, 151
        & &
        \\
        \hline
    \end{tabular}
    \smallskip
    \caption{Primes $\ell$ amenable to 
        Algorithm~\ref{alg:ElkiesWalk} (\algstyle{ElkiesWalk})
        for our candidate isogeny graph, 
        with proposed walk length bounds $M_\ell$.}
    \label{tab:ElkiesSteps}
\end{table}

Using these parameters, we perform one isogeny walk in
approximately 520 seconds.
These timings are~\emph{worst-case}:
the number of isogeny steps is taken
to be exactly $M_\ell$ for each $\ell$.
This is about as fast as Stolbunov's largest parameter~\cite{Stol},
which is for a prime of 428 bits and a keyspace of only 216 bits.

We stress that our implementation
is \emph{not} optimised. General gains in field arithmetic aside,
optimised code could easily beat our proof-of-concept implementation
at critical points of our algorithms, such as the root finding steps 
in Algorithms~\ref{alg:ElkiesFirstStep}
and~\ref{alg:ElkiesNextStep}.

For comparison, without Algorithm~\ref{alg:VéluStep} the
total isogeny walk time would exceed 2000 seconds.
Our ideas thus yield an improvement by a factor of over 4 over the
original protocol. A longer search for efficient public
parameters would bring further improvement.

\section{Conclusion}

We have shown that the Couveignes--Rostovtsev--Stolbunov
framework can be improved to become practical at standard pre- and
post-quantum security levels; 
even more so if an optimized C implementation is made. 
The main obstacle to better performance 
is the difficulty of generating optimal system parameters:
even with a lot of computational power,
we cannot expect to produce ordinary curve parameters
that allow us to use \emph{only} Vélu steps. 
In this regard, the CSIDH protocol~\cite{csidh}, 
which overcomes this problem using
supersingular curves instead of ordinary ones, is promising.

One particularly nice feature of our protocol is its highly efficient key validation,
which opens a lot of cryptographic doors. However, side-channel-resistant
implementations remain an interesting problem for future work.

\bibliographystyle{splncs04}
\bibliography{refs}

\end{document}